\newif\ifarXiv
\arXivtrue   % arXiv
\ifarXiv
\documentclass[12pt]{article}
\usepackage{epsfig,amsmath,amsfonts,amssymb,makeidx,ifthen}
\usepackage[dvips]{color}
\newtheorem{theorem}{Theorem}[section]
\newtheorem{lemma}[theorem]{Lemma}

\makeatletter
\@addtoreset{equation}{section}
\makeatother

\makeatletter
\long\def\@makecaption#1#2{{\small
\advance\leftskip1cm
\advance\rightskip1cm
\vskip\abovecaptionskip
\sbox\@tempboxa{#1: #2}%
\ifdim \wd\@tempboxa >\hsize
 #1: #2\par
\else
\global \@minipagefalse
\hb@xt@\hsize{\hfil\box\@tempboxa\hfil}%
\fi
\vskip\belowcaptionskip}}
\makeatother
\else
\documentclass[smallextended]{svjour3}  % onecolumn (second format)
\usepackage{graphicx,amsmath,amsfonts,amssymb,makeidx,ifthen,color}
\fi
\ifarXiv
\newcommand{\jour}[2]{#1}
\else
\newcommand{\jour}[2]{#2}
\fi
\newcommand{\lb}[1]{\label{e:#1}}
\newcommand{\rlb}[1]{\eqref{e:#1}} 
\newcommand{\nl}{\notag\\}

\newcommand{\sumtwo}[2]%
{\mathop{\sum_{#1}}_{#2}}
\newcommand{\sumthree}[3]%
{\mathop{\mathop{\sum_{#1}}_{#2}}_{#3}}
\newcommand{\sumfour}[4]%
{\mathop{\mathop{\mathop{\sum_{#1}}_{#2}}_{#3}}_{#4}} 
\newcommand{\prodtwo}[2]%
{\mathop{\prod_{#1}}_{#2}}
\newcommand{\mintwo}[2]%
{\mathop{\min_{#1}}_{#2}}
\newcommand{\maxtwo}[2]%
{\mathop{\max_{#1}}_{#2}}
\newcommand{\maxthree}[3]%
{\mathop{\mathop{\max_{#1}}_{#2}}_{#3}}
\newcommand{\limtwo}[2]%
{\mathop{\lim_{#1}}_{#2}}
\newcommand{\suptwo}[2]%
{\mathop{\sup_{#1}}_{#2}}
\newcommand{\supthree}[3]%
{\mathop{\mathop{\sup_{#1}}_{#2}}_{#3}}
\newcommand{\supfour}[4]%
{\mathop{\mathop{\mathop{\sup_{#1}}_{#2}}_{#3}}_{#4}} 
\newcommand{\inftwo}[2]%
{\mathop{\inf_{#1}}_{#2}}
\newcommand{\infthree}[3]%
{\mathop{\mathop{\inf_{#1}}_{#2}}_{#3}}
\newcommand{\inffour}[4]%
{\mathop{\mathop{\mathop{\inf_{#1}}_{#2}}_{#3}}_{#4}} 
\newcommand\calK{{\cal K}}

\newcommand{\bsk}{\boldsymbol{k}}

\newcommand{\bssigma}{\boldsymbol{\sigma}}

\newcommand{\bbC}{\mathbb{C}}

\newcommand{\ep}{\epsilon}
\newcommand{\up}{\uparrow}
\newcommand{\dn}{\downarrow}

\newcommand{\qedm}{\rule{1.5mm}{3mm}}
\newcommand{\cE}{\mathcal{E}}
\newcommand{\bu}{{\bar{u}}}
\newcommand{\bO}{\bar{\mathcal{O}}}
\newcommand{\cOO}{\cO\cup\bO}
\newcommand{\bS}{\hat{\boldsymbol{S}}}
\newcommand{\hS}{\hat{S}}
\newcommand{\cPa}{\mathcal{P}_a}
\newcommand{\hA}{\hat{A}}
\newcommand{\hB}{\hat{B}}

\newcommand{\La}{\Lambda}
\newcommand{\cB}{\mathcal{B}}

\newcommand{\cM}{\mathcal{M}}
\newcommand{\nM}{|\cM|}
\newcommand{\cO}{\mathcal{O}}

\newcommand{\Ham}{\hat{H}}
\newcommand{\hc}{\hat{c}}
\newcommand{\hcd}{\hat{c}^\dagger}
\newcommand{\ha}{\hat{a}}
\newcommand{\had}{\hat{a}^\dagger}
\newcommand{\hb}{\hat{b}}
\newcommand{\hbd}{\hat{b}^\dagger}
\newcommand{\hn}{\hat{n}}
\newcommand{\Smax}{S_\mathrm{max}}
\newcommand{\Stot}{S_\mathrm{tot}}
\newcommand{\hStot}{\hat{S}_\mathrm{tot}}
\newcommand{\bStot}{\hat{\boldsymbol{S}}_\mathrm{tot}}
\newcommand{\ket}[1]{|#1\rangle}
\newcommand{\bra}[1]{\langle#1|}
\newcommand{\vac}{\ket{\Phi_\mathrm{vac}}}
\newcommand{\kPhi}{\ket{\Phi}}
\newcommand{\Ne}{N_\mathrm{e}}
\newcommand{\Nh}{N_\mathrm{h}}
\newcommand{\Cs}{{C,\bssigma}}
\newcommand{\EP}{E_{\Phi}}
\newcommand{\Ef}{E_\mathrm{ferro}}
\newcommand{\kx}{k_\mathrm{x}}
\newcommand{\ky}{k_\mathrm{y}}
\newcommand{\para}[1]{\paragraph{#1}}
\begin{document}
\jour{ %arXiv
\vspace*{1em}
\noindent
{\bf
\Large Metallic ferromagnetism supported by a single band}
\vspace{1mm}
\par\noindent
{\bf\Large in a multi-band Hubbard model
}
\par\bigskip

\noindent
Akinori Tanaka\footnote{
Department of General Education,
Ariake National College of Technology, Omuta, 
Fukuoka 836-8585, Japan
} and Hal Tasaki\footnote{
Department of Physics, Gakushuin University, Mejiro, Toshima-ku, 
Tokyo 171-8588, Japan
}

}
{
% journal 

\title{Metallic ferromagnetism supported by a single band
in a multi-band Hubbard model}

\titlerunning{Metallic ferromagnetism in a multi-band Hubbard model}

\author{Akinori Tanaka  \and
        Hal Tasaki     
}


\institute{A. Tanaka \at
              Department of General Education,
	      Ariake National College of Technology, Omuta, 
	      Fukuoka 836-8585, Japan \\
              \email{akinori@ariake-nct.ac.jp}           %  \\
           \and
           H. Tasaki \at
	      Department of Physics, Gakushuin University, Mejiro, Toshima-ku, 
	      Tokyo 171-8588, Japan\\
	      \email{hal.tasaki@gakushuin.ac.jp}
}

\date{Received: date / Accepted: date}

\maketitle
} 
%
%
%%%%%%%%%%%%%%%%%%%%%%%%%%%%%%%%%%%%%%%
\begin{abstract}
We construct a multi-band Hubbard model on the lattice obtained by ``decorating'' a closely packed $d$-dimensional lattice $\cM$ (such as the triangular lattice) where $d\ge2$.
We take the limits in which the Coulomb interaction and the band gap become infinitely large.
Then there remains only a single band with finite energy, on which electrons are supported.
Let the electron number be $\Ne=\nM-\Nh$, where $\nM$ corresponds to the electron number which makes the lowest (finite energy) band half-filled, and $\Nh$ is the number of ``holes''.
It is expected that the model exhibits metallic ferromagnetism if $\Nh/\nM$ is nonvanishing but sufficiently small.
We prove that the ground states exhibit saturated ferromagnetism if $\Nh\le(\text{const.})\nM^{2/(d+2)}$, and exhibit (not necessarily saturated) ferromagnetism if  $\Nh\le(\text{const.})\nM^{(d+1)/(d+2)}$.
This may be regarded as a rigorous example of metallic ferromagnetism provided that the system size $\nM$ is not too large.
\jour{}
{
}
\end{abstract}
%
%%%%%%%%%%%%%%%%%%%%%%%%%%%%%%%%%%%%%%%
%%%%%%%%%%%%%%%%%%%%%%%%%%%%%%%%%%%%%%%
%
\jour{%arXiv
\if0
\makeatletter
\renewcommand*{\l@section}
{\@dottedtocline{1}{0pt}{1.5em}}
\makeatother
\fi
\tableofcontents
}
{}
%%%%%%%%%%%%%%%%%%%%%%%%%%%%%%%%%%%%%%
%%%%%%%%%%%%%%%%%%%%%%%%%%%%%%%%%%%%%%
\section{Introduction}

Since the pioneering work of Heisenberg \cite{Heisenberg}, it is understood that ferromagnetism observed in nature is generated by quantum many-body effects of many fermions and the Coulomb interaction between electrons.
It is a challenging theoretical  problem to confirm this scenario by showing that only short-range hopping of electrons and the spin-independent Coulomb interaction can lead to ferromagnetism in the concrete setting of the Hubbard model.
See \cite{Lieb95,Tasaki98a,Tasaki98b} for early reviews.

There have been a number of  rigorous examples of ferromagnetism (or ferrimagnetism \cite{Lieb}) in the Hubbard model, and it is clear by now that certain versions of the model do generate ferromagnetism. 
Important examples include special classes of the Hubbard model with a dispersionless band (flat-band ferromagnetism) introduced by Mielke \cite{Mielke92,Mielke93} and by Tasaki \cite{Tasaki92,MielkeTasaki}, and some models obtained by modifying the flat-band models \cite{Tasaki94,Tasaki95,Tasaki96,TanakaUeda,Tasaki03}.
The latter examples are of special importance since they provide rigorous examples of itinerant ferromagnetism in many-electron models without any singularities.

A common feature of all these rigorous examples of ferromagnetism is that they describe insulators.
Metallic ferromagnetism, in which same electrons contribute both to magnetism and conduction, is clearly more interesting and challenging to understand.
The older example by Thouless \cite{Thouless} and Nagaoka \cite{Nagaoka} may not be regarded as metallic ferromagnetism since it allows only a single carrier in the whole system.

As far as we know the first rigorous example of metallic ferromagnetism
in the Hubbard model was presented by Tanaka and Idogaki \cite{TanakaIdogaki99}, who treated a
quasi one-dimensional model.
See also \cite{Kubo} for earlier work which employed similar ideas in a different model.
But the physics of one-dimensional electron systems is rather special, and the basic mechanism is captured by the Perron-Frobenius argument \cite{Tasaki98a}.
See also \cite{LiLiebWu2014} where a class of multi-orbital Hubbard models in higher dimensions with restricted hopping is treated\footnote{
In these models, the motion of electrons in each orbital is restricted to one direction.
This construction makes the Perron-Frobenius theorem applicable.}.
The same argument never works for models in genuine two and higher dimensions.

In \cite{TanakaTasaki2007}, Tanaka and Tasaki presented the first rigorous example of metallic ferromagnetism in the Hubbard model in two and higher dimensions.
In this work, a multi-band Hubbard model with short-range (but admittedly complicated) hoppings was constructed, and it was proved that the ground state of the model exhibits saturated ferromagnetism in the limits where the Coulomb interaction $U$ and the band gap tend to infinity.
In the ground state, electrons are supported by the two lowest bands, where the lowest band is half-filled (which effectively means full-filling for fully polarized states) and the second lowest band carries less electrons than the half-filling.
The ground state is metallic since the second lowest band can carry conduction.

%%%%%%%%%%%%%%%%%%%%%%%%%%%%%%%%%%%%%%
\begin{figure}[btp]
\begin{center}
\jour{\includegraphics[width=8cm]{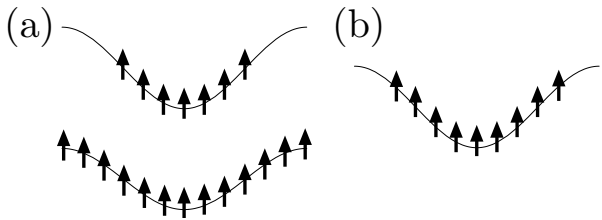}\vspace{-5mm}}
{\includegraphics[width=7cm]{bands.eps}}
\end{center}
\caption[dummy]{
Schematic pictures of the band structures in the previous \cite{TanakaTasaki2007} and the present model.
(a)~In the previous model, ferromagnetism was supported by the two lowest bands, where the lower one was half-filled.
(b)~In the present model, ferromagnetism is supported by the lowest band alone, which is nearly half-filled.
}
\label{fig:bands}
\end{figure}
%%%%%%%%%%%%%%%%%%%%%%%%%%%%%%%%%%%%%%

Although the model of \cite{TanakaTasaki2007} certainly captures some essence of metallic ferromagnetism, it is unsatisfactory in the sense that the ferromagnetism is supported by two bands.
One may interpret that the lowest band is exhibiting ferromagnetism from the mechanism similar to that in the insulating ferromagnets discussed above.
See Fig.~\ref{fig:bands}.

It is thus desirable to have examples of Hubbard models where metallic ferromagnetism takes place only within a single (conduction) band.
Then we should have ferromagnetism generated by a mechanism intrinsic to metallic systems.
This is the goal of the present study, which we partially achieve.

We shall construct a new class of Hubbard model on lattices obtained by decorating 
closely packed lattices in two or higher dimension (such as the triangular lattice).
The model has short-ranged but somewhat complicated hoppings, and the on-site Coulomb interaction.
We shall take the limits where the Coulomb interaction $U$ and the band gap become infinitely large.
Then there remains only a single band with finite energy, on which electrons are supported.
We shall prove that the ground state of the model exhibits ferromagnetism for certain ranges of the electron number.
Although we are not able to treat the case where the density of carrier remains nonzero in the infinite volume limit, our result may be regarded as examples of metallic ferromagnetism when the system size is not too large.

The present work is also of interest from technical point of view since we here develop some new techniques for dealing with metallic ferromagnetism.
In particular we shall prove Theorems~\ref{t:F} and \ref{t:Ft} about the ferromagnetism in ground states without constructing the ground states explicitly.
This should have clear advantage over the methods in our previous works, where (fully polarized) ground states were always constructed explicitly.
The most important argument, which is presented in section~\ref{s:Ft}, enables us to relate the kinetic energy with the total spin in a many-electron state.

In the present paper, we focus on models in dimensions two or higher, and do not discuss one dimensional systems.

\bigskip

The present paper is organized as follows.

In section~\ref{s:model}, we shall describe our model in general setting, and state main results about metallic ferromagnetism for the model based on a general $d$-dimensional closely packed lattice.

The remaining sections are devoted to the proof.
In section~\ref{s:tJ}, by following the method developed in our earlier works (see, e.g., \cite{Tasaki98a}), we show that 
strong ferromagnetic coupling between electrons in the lowest band is generated.
The rest of the proof employs new techniques developed for the present work.
After proving an essential lower bound on the energy in section~\ref{s:E}, we concentrate on the model based on the triangular lattice, and prove the theorem about saturated ferromagnetism in section~\ref{s:SFt}, and that about (not necessarily saturated) ferromagnetism in section~\ref{s:Ft}.
Finally, in section~\ref{s:lattice}, we describe the extension to general lattices.

\section{Model and main results}
\label{s:model}
We shall describe our model in general, and state main theorems.

\para{Lattice}
Let $\cM$ be a finite lattice, or, more precisely, a set of sites, whose elements are denoted as $x,y\ldots\in\cM$.
A bond $(x,y)$ is an ordered pair of distinct sites $x,y\in\cM$ which are regarded to be neighboring with each other.
We denote by $\cB$ 
the set of all bonds.
We assume that, for any $x,y\in\cM$, at most one of $(x,y)$ or $(y,x)$ belongs to $\cB$.
The whole lattice $\cM$ is assumed to be connected via bonds in $\cB$.
We finally assume that the coordination number of the lattice is uniform and is equal to $\zeta$, i.e., for any $x\in\cM$ there are exactly $\zeta$ sites $y\in\cM$ such that $(x,y)\in\cE$,
where $\cE$ is defined as
$\cE:=\cB\cup\{(y,x)\,|\,(x,y)\in\cB\}$
\footnote{
We write $A:=B$ or $B=:A$ when $A$ is defined in terms of $B$.
}.

\begin{figure}[btp]
\begin{center}
\jour{\includegraphics[width=7cm]{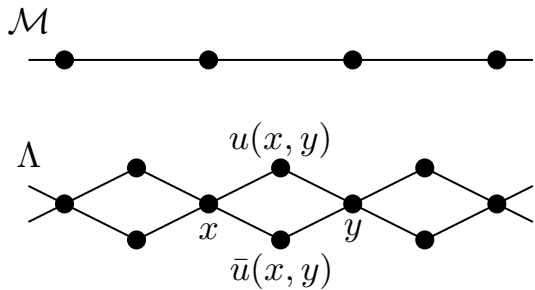}\vspace{-5mm}}
{\includegraphics[width=6cm]{1d.eps}}
\end{center}
\caption[dummy]{
The lattice structure when $\cM$ is a one-dimensional chain.
Although we only study lattices in two or higher dimensions, this illustrates the structure of the lattice.
}
\label{fig:1d}
\end{figure}

With each (ordered) bond $(x,y)\in\cB$, we associate two additional sites $u$ and $\bu$, or, more precisely, $u(x,y)$ and $\bu(x,y)$.
We denote by $\cO$ and $\bO$ the collections of $u(x,y)$ and $\bu(x,y)$, 
respectively, for all $(x,y)\in\cB$.
We define the Hubbard model on the decorated lattice $\La=\cM\cup\cOO$.
See Figure~\ref{fig:1d}.
Similar lattices were studied in \cite{TanakaIdogaki2001,Sekizawa}.

%%%%%%%%%%%%%%%%%%%%%%%%%%%%%%%%%%%%%%
%%%%%%%%%%%%%%%%%%%%%%%%%%%%%%%%%%%%%%

\para{Fermion operators}
For each site $z\in\La$ and spin index $\sigma=\up,\dn$, we denote by $\hc_{z,\sigma}$ the standard fermion operator which annihilates an electron at site $z$ with spin $\sigma$.
The corresponding creation operator and the number operator are $\hcd_{z,\sigma}$ and $\hn_{z,\sigma}:=\hcd_{z,\sigma}\hc_{z,\sigma}$, respectively.
We denote by $\vac$ the unique normalized state with no electrons on the system.
We consider the Hilbert space with a fixed electron number $\Ne$, which is assumed to satisfy\footnote{
We denote by $|S|$ the number of elements in a finite set $S$.
} $\Ne\le\nM$.

As usual we define the spin operator 
$\bS_z=(\hS^{(1)}_z,\hS^{(2)}_z,\hS^{(3)}_z)$ at site $z$ 
by $\hS^{(1)}_z=(\hS^+_z+\hS^-_z)/2$, $\hS^{(2)}_z=(\hS^+_z-\hS^-_z)/(2i)$ with
\begin{equation}
\hS^+_z:=\hcd_{z,\up}\hc_{z,\dn},\quad
\hS^-_z:=\hcd_{z,\dn}\hc_{z,\up},\quad
\hS^{(3)}_z:=\frac{1}{2}(\hn_{z,\up}-\hn_{z,\dn}),
\lb{Sops}
\end{equation}
and also define the total spin operator $\bStot=(\hStot^{(1)},\hStot^{(2)},\hStot^{(3)})$ 
by $\sum_{z\in\La}\hS_z^{(l)}$ with $l=1,2,3$.
The eigenvalue of $(\bStot)^2$ is denoted as $\Stot(\Stot+1)$, where $\Stot=\Smax,\Smax-1,\ldots,0$ or $1/2$.
We have defined the maximum spin as $\Smax:=\Ne/2$.

\begin{figure}[btp]
\begin{center}
\jour{\includegraphics[width=9cm]{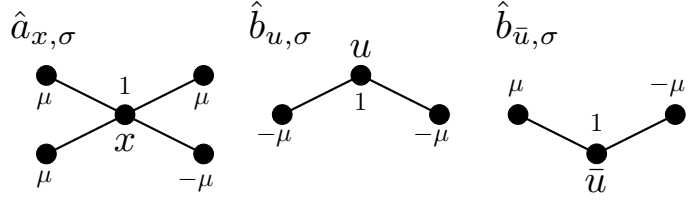}\vspace{-7mm}}
{\includegraphics[width=8cm]{ab.eps}}
\end{center}
\caption[dummy]{
The states corresponding to the operators $\ha_{x,\sigma}$, $\hb_{u,\sigma}$, and $\hb_{\bu,\sigma}$ for the one dimensional model on the lattice in Figure~\ref{fig:1d}.
}
\label{fig:ab}
\end{figure}

To describe our model, we define special fermion operators.
Fix a constant $\mu>0$.
For each $x\in\cM$ and $\sigma=\up,\dn$, let 
\jour{
\begin{equation}
\ha_{x,\sigma}:=\frac{1}{\sqrt{1+2\zeta\mu^2}}
\Bigl\{\hc_{x,\sigma}
+\mu\!\!\!\!\sumtwo{y\in\cM}{((x,y)\in\cB)}\!\!\!\!
(\hc_{u(x,y),\sigma}-\hc_{\bu(x,y),\sigma})
+\mu\!\!\!\!\sumtwo{y\in\cM}{((y,x)\in\cB)}\!\!\!\!
(\hc_{u(y,x),\sigma}+\hc_{\bu(y,x),\sigma})
\Bigr\}.
\lb{adef}
\end{equation}
}
{
\begin{eqnarray}
\ha_{x,\sigma}&:=&\frac{1}{\sqrt{1+2\zeta\mu^2}}
\Bigl\{\hc_{x,\sigma}
+\mu\!\!\!\!\sumtwo{y\in\cM}{((x,y)\in\cB)}\!\!\!\!
(\hc_{u(x,y),\sigma}-\hc_{\bu(x,y),\sigma})
\nonumber\\
&&\hspace*{4cm}
+\mu\!\!\!\!\sumtwo{y\in\cM}{((y,x)\in\cB)}\!\!\!\!
(\hc_{u(y,x),\sigma}+\hc_{\bu(y,x),\sigma})
\Bigr\}.
\lb{adef}
\end{eqnarray}
}
For each $u\in\cO$ and $\sigma$, we let
\begin{equation}
\hb_{u,\sigma}:=\hc_{u,\sigma}-\mu(\hc_{x,\sigma}+\hc_{y,\sigma}),
\lb{bdef1}
\end{equation}
where $(x,y)\in\cB$ is the unique
bond such that $u=u(x,y)$, and similarly for each $\bu\in\bO$ and $\sigma$,
\begin{equation}
\hb_{\bu,\sigma}:=\hc_{\bu,\sigma}
+
\mu(\hc_{x,\sigma}-\hc_{y,\sigma}),
\lb{bdef2}
\end{equation}
where $(x,y)\in\cB$ is the unique bond such that $\bu=\bu(x,y)$.
See Figure~\ref{fig:ab}.

%%%%%%%%%%%%%%%%%%%%%%%%%%%%%%%%%%%%%%
%%%%%%%%%%%%%%%%%%%%%%%%%%%%%%%%%%%%%%

It is easily verified that 
the $\ha$-operators satisfy the standard anticommutation relations
\begin{equation}
\{\had_{x,\sigma},\ha_{y,\tau}\}=\delta_{x,y}\delta_{\sigma,\tau},
\lb{aa}
\end{equation}
for any $x,y\in\cM$ and $\sigma,\tau=\up,\dn$.
It also holds that
\begin{equation}
\{\had_{x,\sigma},\hb_{v,\tau}\}=0,
\lb{ab}
\end{equation}
for any $x\in\cM$, $v\in\cOO$, and $\sigma,\tau=\up,\dn$.
These relations together with $\{\hc_{v,\sigma}^\dagger,\hb_{w,\tau}\}=\delta_{v,w}\delta_{\sigma,\tau}$
for $v,w\in\cO\cup\bO$ and $\sigma,\tau=\up,\dn$
also imply that the single electron states $\ha_{x,\sigma}^\dagger\vac$ and $\hb_{v,\sigma}^\dagger\vac$ are linearly independent,
i.e.,
any single electron state on $\La$ can be represented by a suitable linear combination 
of $\ha_{x,\sigma}$ with $x\in\cM$ and $\hb_{v,\sigma}$ with $v\in\cOO$.

\para{The model}
We study the Hubbard model with the Hamiltonian
\begin{equation}
\Ham:=t\sumtwo{(x,y)\in\cE}{\sigma=\up,\dn}\had_{x,\sigma}\ha_{y,\sigma}
+s\sumtwo{v\in\cOO}{\sigma=\up,\dn}\hbd_{v,\sigma}\hb_{v,\sigma}
+U\sum_{z\in\La}\hn_{z,\up}\hn_{z,\dn},
\lb{Ham}
\end{equation}
where $s$, $t$, and $U$ are positive parameters.

Note that, by using the definitions \rlb{adef}, \rlb{bdef1}, and \rlb{bdef2}, one can rewrite the first two sums in \rlb{Ham} in the standard form $\sum_{z,z'\in\La,\sigma=\up,\dn}t_{z,z'}\,\hcd_{z,\sigma}\hc_{z'\sigma}$.
The hopping amplitude $t_{z,z'}$ is admittedly complicated, but is short ranged.

Consider a single electron state of the form $\ket{\Psi}=\sum_{x\in\cM}\psi_x\,\had_{x,\up}\vac$, where $\psi_x\in\bbC$ are coefficients.
By using the fact that $\hc_{z,\sigma}\vac=0$ for any $z\in\La$ and $\sigma$, and the anticommutation relations \rlb{aa} and \rlb{ab}, one finds that
 the Schr\"odinger equation $\ep\ket{\Psi}=\Ham\ket{\Psi}$ becomes
\begin{equation}
\ep\,\psi_x=t\!\!\sumtwo{y\in\cM}{((x,y)\in\cE)}\!\!\psi_y\quad\text{for any $x\in\cM$},
\lb{sSch}
\end{equation}
which is the simplest tight-binding Schr\"odinger equation on the lattice $\cM$.
We shall refer to the energy band formed by the solutions of \rlb{sSch} as the $a$-band.
Similarly, by solving the Schr\"odinger equation $\ep\ket{\Psi}=\Ham\ket{\Psi}$
with $\ket{\Psi}=\sum_{v\in\cO\cup\bO}\psi_v\,\hbd_{v,\up}\vac$, we find other bands, to which we shall refer as the $b$-bands.
With some calculations, one finds that the $b$-bands consist of two flat-bands, 
one at energy $s$ with $(\zeta-1)|\cM|$-fold degeneracy, 
and the other at energy $(1+2\zeta\mu^2)s$ with $|\cM|$-fold degeneracy.  
Since we shall let $s\up\infty$, electrons are always supported by the $a$-band.

\para{Main results}
Throughout the present paper we fix the electron number $\Ne\le\nM$, and consider the limits $U\up\infty$ and  $s\up\infty$.
While the first limit makes the Coulomb repulsion infinitely strong, the latter infinitely lifts the energy of bands other than the $a$-band.
Although it is not desirable to take such singular limits, this seems necessary for the moment to control the model rigorously.

Let us describe our theorems for the general class of models.

We first only assume that the lattice $\cM$ is connected and has a uniform coordination number $\zeta$.
Note that the model exhibits flat-band ferromagnetism when $t=0, U>0$ and $\Ne=|\cM|$.
The following corresponds to our older results for the flat-band ferromagnetism.
%%%%%%%%%%%%%%%%%%%%%%
\begin{theorem}\label{t:Feasy0}
Suppose that the electron number is $\Ne=\nM$.
Then, in the limits $U\up\infty$ and $s\up\infty$,  ground states of the model have $\Stot=\Smax$ and are unique up to the trivial $(2\Smax+1)$-fold degeneracy.
\end{theorem}
%%%%%%%%%%%%%%%%%%%%%%
This theorem is proved in Section~\ref{s:tJ} as a straightforward consequence of the strong ferromagnetic coupling in the lowest band.
Since $\Smax:=\Ne/2$ is the maximum possible total spin, the above theorem establishes that the ground states exhibit saturated ferromagnetism\footnote{\label{footnote:modela}
For models obtained by adding suitable $\ha_{x,\sigma}^\dagger \ha_{x,\sigma}$ terms to $\Ham$, 
it is possible to prove the same statement   
for sufficiently large $U$ and $s$.
The proof uses techniques similar to that in \cite{Tasaki94,Tasaki95,Tasaki96,TanakaUeda,Tasaki03}.
}.
Although the electron number $\Ne=\nM$ corresponds to the half-filing of the lowest $a$-band, the band is effectively fully filled because the state has the maximum spin.
It is strongly expected that the ground states in this case are {\em ferromagnetic Mott insulator}\/.

If we lower the electron number from $\nM$, movable ``holes'' are doped into the $a$-band.
It is expected that the ground states with sufficiently many holes are conducting, i.e., exhibit metallic ferromagnetism.
It should be noted however that the electron number cannot be too small to maintain global ferromagnetism.
In fact it is expected that, in the dimensions two or higher, the present model becomes a paramagnetic metal 
at low electron density
even though strong ferromagnetic coupling is generated. See section~\ref{s:tJ}.

The following is an easy result which works when the number of holes is less than the coordination number $\zeta$ of the lattice $\cM$.
%%%%%%%%%%%%%%%%%%%%%%
\begin{theorem}\label{t:Feasy1}
Suppose that the electron number $\Ne$ satisfies $\nM\ge\Ne>\nM-\zeta$.
Then, in the limits $U\up\infty$ and $s\up\infty$, ground states of the model have $\Stot=\Smax$.
\end{theorem}
%%%%%%%%%%%%%%%%%%%%%%
The theorem is proved in Section~\ref{s:tJ} again as a straightforward consequence of the strong ferromagnetic coupling in the lowest band.

In order to increase the number of holes, we need to impose some conditions on the lattice.
In the following we assume that the lattice $\cM$ has dimension $d\ge2$ and belongs to a class which we call ``closely packed lattices''.
The precise definition of closely packed lattices is given in section~\ref{s:lattice}.
We here note that examples include the triangular lattice with $d=2$, and the checkerboard lattice\footnote{
The checkerboard lattice is a $d$-dimensional extension of the face centered cubic lattice,
whose lattice points are given by $(\mathrm{x}_1,\mathrm{x}_1,\dots,\mathrm{x}_d)$
with integers $\mathrm{x}_l$ such that $\sum_{l=1}^{d}\mathrm{x}_l=\mathrm{even}$.
While the triangular lattice and the checkerboard lattices with $d=3,4,5$ provide the densest possible lattice packing, the checkerboard lattices with $d\ge6$ do not.
Thus our notion of ``closely packed lattices'' is different from the densest packing.
} with $d\ge3$.
%%%%%%%%%%%%%%%%%%%%%%
\begin{theorem}\label{t:SF}
Suppose that the electron number $\Ne$ satisfies $\nM\ge\Ne\ge\nM-C\,\nM^{2/(d+2)}$, where $C$ is a constant depending only on the structure (and not on the size) of the lattice.
Then, in the limits $U\up\infty$ and $s\up\infty$, ground states of the model have $\Stot=\Smax$.
\end{theorem}
%%%%%%%%%%%%%%%%%%%%%%
The theorem is proved for the triangular lattice in section~\ref{s:SFt}. The extension to other lattices is discussed in section~\ref{s:lattice}.
This theorem again establishes that the ground states exhibit saturated ferromagnetism.

If we relax the condition for ferromagnetism, we can considerably increase the number of holes as in the following theorem, which is the most important result of the present paper.
%%%%%%%%%%%%%%%%%%%%%%
\begin{theorem}\label{t:F}
Fix an arbitrary constant $\nu$ such that $0<\nu<1$.
Suppose that the electron number $\Ne$ satisfies $\nM\ge\Ne\ge\nM-C'\,(1-\nu)^{(d-1)/(d+2)}\nM^{(d+1)/(d+2)}$, where $C'$ is a constant depending only on the structure (and not on the size) of the lattice.
Then, in the limits $U\up\infty$ and $s\up\infty$, ground states of the model have $\Stot\ge\nu\Smax$.
\end{theorem}
%%%%%%%%%%%%%%%%%%%%%%
The theorem is proved for the triangular lattice in section~\ref{s:Ft}, and the extension to other lattices is discussed in section~\ref{s:lattice}.

Note that the inequality $\Stot\ge\nu\Smax$ implies that a ground state has a magnetic moment proportional to the system size, and hence is ferromagnetic\footnote{
Although the inequality $\Stot\ge\nu\Smax$ suggests the possibility that the ground state exhibits partial ferromagnetism where $\Stot/\Smax$ is strictly less than one (even in the thermodynamic limit), we suspect that the ground state indeed has $\Stot=\Smax$.
}.
In this case the maximum possible number of holes that we can introduce to the $a$-band is $\Nh=C'\,(1-\nu)^{(d-1)/(d+2)}\nM^{(d+1)/(d+2)}$.
Although the hole number $\Nh$ diverges as $\nM\up\infty$, the density of the holes $\Nh/\nM=C'\,(1-\nu)^{(d-1)/(d+2)}\nM^{-1/(d+2)}$, unfortunately, converges to zero.
It is of course most desirable to allow a finite density of holes, i.e., to prove the same result for $\nM\ge\Ne\ge\nM-\mathrm{const}\nM$, but we are for the moment far from showing such results.
Note however that, as the dimension $d$ gets large, $\nM^{(d+1)/(d+2)}$ becomes close to\footnote{
This is a big advantage of Theorem~\ref{t:F} over Theorem~\ref{t:SF}.
Note that $\nM^{2/(d+2)}$ that appears in Theorem~\ref{t:SF} converges to zero as $d\up\infty$.
} $\nM$.
This fact is also consistent with our intuition that ferromagnetism is more stable in higher dimensions.

We may conclude that Theorem~\ref{t:F} shows the existence of metallic ferromagnetism in the ground state of our model for not too large $\nM$, where we have a sufficient density of holes to carry electric current.
Unfortunately the conclusion does not carry over to the thermodynamic limit $\nM\up\infty$, where the hole density vanishes.
It is most desirable to find a method for treating the situation with a finite density of holes.

%%%%%%%%%%%%%%%%%%%%%%%%%%%%%%%%%%%%%%%%%%
\section{Ferromagnetic coupling in the lowest band}
\label{s:tJ}
We shall prove the theorems in the following four sections.

Here we treat the model on a general connected lattice with any electron number $\Ne\le\nM$.
By employing the techniques developed for insulating ferromagnets (see, e.g., \cite{Tasaki98a}), we show that strong ferromagnetic coupling is generated in a state which has a finite energy in the limits $U\up\infty$ and $s\up\infty$.
One can also say that we show the equivalence of our model to the ferromagnetic $t$-$J$ model with $J\up\infty$ defined on the lattice $\cM$.
Although one might imagine that the existence of strong ferromagnetic coupling is enough to establish the existence of ferromagnetism, this is indeed far from the case, as we shall explain below.

\para{Basic constraints for finite energy states}
To begin with, note that any state $\kPhi$ with the electron number $\Ne\le\nM$ is written as
\jour{
\begin{equation}
\kPhi=\sumtwo{C_\up,C_\dn\subset\cM}{C'_\up,C'_\dn\subset\cOO}
\psi(C_\up,C_\dn,C'_\up,C'_\dn)\,
\Bigl(\prod_{x\in C_\up}\had_{x,\up}\Bigr)
\Bigl(\prod_{x\in C_\dn}\had_{x,\dn}\Bigr)
\Bigl(\prod_{v\in C'_\up}\hbd_{v,\up}\Bigr)
\Bigl(\prod_{v\in C'_\dn}\hbd_{v,\dn}\Bigr)
\vac,
\lb{Phiaabb}
\end{equation}}
{\begin{eqnarray}
\kPhi&=&\sumtwo{C_\up,C_\dn\subset\cM}{C'_\up,C'_\dn\subset\cOO}
\psi(C_\up,C_\dn,C'_\up,C'_\dn)
\nonumber\\
&&
\hspace*{1.5cm}
\times
\Bigl(\prod_{x\in C_\up}\had_{x,\up}\Bigr)
\Bigl(\prod_{x\in C_\dn}\had_{x,\dn}\Bigr)
\Bigl(\prod_{v\in C'_\up}\hbd_{v,\up}\Bigr)
\Bigl(\prod_{v\in C'_\dn}\hbd_{v,\dn}\Bigr)
\vac,
\lb{Phiaabb}
\end{eqnarray}}
where the sums are taken over all the subsets $C_\up$, $C_\dn$, $C'_\up$, and $C'_\dn$ such that $|C_\up|+|C_\dn|+|C'_\up|+|C'_\dn|=\Ne$, and $\psi(C_\up,C_\dn,C'_\up,C'_\dn)\in\bbC$ are arbitrary coefficients.
Here, and throughout the present paper, we assume that there are fixed (but arbitrary) ordering in the elements of $\cM$, $\cO$, and $\bO$, and the products are arranged according to the ordering.

Let us examine necessary and sufficient conditions for a state $\kPhi$ to have a finite energy in the limits $U\up\infty$ and $s\up\infty$, i.e.,
\begin{equation}
\lim_{s\up\infty}\lim_{U\up\infty}\bra{\Phi}\Ham\kPhi<\infty.
\lb{fe}
\end{equation}
Note that $\kPhi$ is regarded here as a given state, which does not vary with $s$ or $U$.
We shall refer to a state $\kPhi$ satisfying the condition \rlb{fe} as a finite energy state.

It may be obvious, 
and can be shown (see, e.g., section 8.2 of \cite{Tasaki98b}) that the condition \rlb{fe} is satisfied if and only if
\begin{equation}
\sumtwo{v\in\cOO}{\sigma=\up,\dn}\hbd_{v,\sigma}\hb_{v,\sigma}\kPhi=0,
\lb{fe2}
\end{equation}
and
\begin{equation}
\sum_{z\in\La}\hn_{z,\up}\hn_{z,\dn}\kPhi=0.
\lb{ccP}
\end{equation}

%%%%%%%%%%%%%%%%%%%%%%%%%%%%%
\para{Effective interaction in the $a$-band}
By noting that $\hbd_{v,\sigma}\hb_{v,\sigma}\ge0$, 
the condition \rlb{fe2} further reduces to  detailed conditions
\begin{equation}
\hb_{v,\sigma}\kPhi=0\quad\text{for every $v\in\cOO$ and $\sigma=\up,\dn$}.
\lb{bP}
\end{equation}
From \rlb{bP}, one finds that $\kPhi$ cannot contain any $\hb$-states.
In other words one has $\psi(C_\up,C_\dn,C'_\up,C'_\dn)=0$ in the expansion \rlb{Phiaabb} unless $C'_\up=C'_\dn=\emptyset$.
Therefore, a finite energy state must be expanded by using the states
\begin{equation}
|\Phi(C_\up,C_\dn)\rangle=
\Bigl(\prod_{x\in C_\up}\had_{x,\up}\Bigr)
\Bigl(\prod_{x\in C_\dn}\had_{x,\dn}\Bigr)
\vac,
\end{equation}
with $C_\up,C_\dn\subset\cM$.
We denote by $\cPa$ the projection operator onto the Hilbert space spanned by these states.

In order to make use of the condition \rlb{ccP}, it is useful to consider the effective interaction
\begin{equation}
 \hat{H}_\mathrm{eff} := \cPa\left(U\sum_{z\in\La} \hn_{z,\up}\hn_{z,\dn}\right)\cPa,
\lb{Heffdef}
\end{equation}
which is the on-site Coulomb interaction restricted to the $a$-band.
In what follows we shall represent $\hat{H}_\mathrm{eff}$ in terms of the $\ha$-operators.

First we consider the on-site interaction $\hn_{x,\up}\hn_{x,\dn}=(\hc_{x,\dn}\hc_{x,\up})^\dagger(\hc_{x,\dn}\hc_{x,\up})$
at the site $x\in\cM$. 
To see how $\hc_{x,\sigma}$ acts on $|\Phi(C_\up,C_\dn)\rangle$, we expand it as 
\begin{equation}
 \hc_{x,\sigma}=\sum_{y\in\cM}\alpha_{y,x} \ha_{y,\sigma}+\sum_{v\in\cOO}\beta_{v,x}\hb_{v,\sigma}. 
 \lb{cexpansion}
\end{equation}
By recalling \rlb{adef}, \rlb{aa} and \rlb{ab}, we find that the coefficients of the $\ha$-operators are given 
by $\alpha_{y,x}=\{\ha_{y,\sigma}^\dagger,\hc_{x,\sigma}\}=(1+2\zeta\mu^2)^{-1/2}\delta_{x,y}$.
This implies 
\begin{equation}
 \hc_{x,\dn}\hc_{x,\up}|\Phi(C_\up,C_\dn)\rangle=\frac{1}{1+2\zeta\mu^2}\ha_{x,\dn}\ha_{x,\up}|\Phi(C_\up,C_\dn)\rangle ,
\end{equation}
where we noted that $\hb_{v,\sigma}|\Phi(C_\up,C_\dn)\rangle=0$.
We thus obtain
\begin{equation}
 \cPa U\hn_{x,\up}\hn_{x,\dn} \cPa = \cPa U^\prime \hn_{x,\up}^{a}\hn_{x,\dn}^{a} \cPa,
\end{equation}
where $U^\prime=U/(1+2\zeta\mu^2)^2$ and $\hn_{x,\sigma}^{a}:=\ha_{x,\sigma}^\dagger \ha_{x,\sigma}$.

Next let us consider the on-site interactions at the sites $u=u(x,y)$ and $\bu=\bu(x,y)$ 
corresponding to the bond $(x,y)\in\cB$.
We expand $\hc_{u,\sigma}$ and $\hc_{\bu,\sigma}$ exactly as in \rlb{cexpansion} to get
\begin{gather}
\hc_{u,\sigma}=\frac{\mu}{\sqrt{1+2\zeta\mu^2}}(\ha_{x,\sigma}+\ha_{y,\sigma})+(\text{$\hb$-operators}),\\
\hc_{\bu,\sigma}=\frac{\mu}{\sqrt{1+2\zeta\mu^2}}(-\ha_{x,\sigma}+\ha_{y,\sigma})+(\text{$\hb$-operators}).
\end{gather}
By using these expansions, one finds (after a straightforward calculation) that 
\begin{equation}
 \cPa \bigl\{U (\hn_{u,\up}\hn_{u,\dn}+\hn_{\bu,\up}\hn_{\bu,\dn})\bigr\} \cPa 
 =
 \cPa \bigl\{W (\hA_{x,y}^\dagger\,\hA_{x,y}+\hB_{x,y}^\dagger\,\hB_{x,y} )\bigr\}\cPa,
\end{equation}
where $W=2U\mu^4/(1+2\zeta\mu^2)^2$, and
\begin{gather}
  \hA_{x,y}:=\ha_{x,\dn}\ha_{x,\up} + \ha_{y,\dn}\ha_{y,\up} \,,\\ 
  \hB_{x,y}:=\ha_{x,\dn}\ha_{y,\up} + \ha_{y,\dn}\ha_{x,\up} \,,
  \lb{Bdef}
\end{gather}
for $(x,y)\in\cB$.

We have thus found that a state $\kPhi$ satisfies the finite energy condition \rlb{fe} if and only if it is expanded as
\begin{equation}
 \kPhi = \sum_{C_\up,C_\dn\subset \cM}\psi(C_\up,C_\dn)|\Phi(C_\up,C_\dn)\rangle\ ,
\lb{exp1}
\end{equation} 
with coefficients $\psi(C_\up,C_\dn)\in\bbC$, and satisfies 
\begin{equation}
\hat{H}_\mathrm{eff}\kPhi=0.
\lb{Heff0}
\end{equation}
The condition \rlb{exp1} follows from \rlb{bP}, and the condition \rlb{Heff0} is nothing but \rlb{ccP}.
The effective interaction \rlb{Heffdef} is expressed as
\begin{equation}
\hat{H}_\mathrm{eff}
=\cPa\Bigl\{
U^\prime\sum_{x\in\cM} \hn_{x,\up}^{a}\hn_{x,\dn}^{a}+W\sum_{(x,y)\in\cB}(\hA_{x,y}^\dagger\,\hA_{x,y}+\hB_{x,y}^\dagger\,\hB_{x,y})\Bigr\}\cPa.
\lb{Heff}
\end{equation}

It is essential to observe that the term $\hB_{x,y}^\dagger \hB_{x,y}$ in \rlb{Heff} describes the ferromagnetic exchange interaction between two electrons in the $\ha_x$ and $\ha_y$ states.
This is explicitly seen by rewriting it as
\begin{equation}
 \hB_{x,y}^\dagger \hB_{x,y}=-2\left(\hat{\boldsymbol{S}}_x^a\cdot\hat{\boldsymbol{S}}_y^a
                          -\frac{\hn_x^a\hn_y^a}{4}\right),
\lb{eff_ferro_int}
\end{equation}
where $\hn_{z}^{a}:=\hn_{z,\up}^{a}+\hn_{z,\dn}^{a}$
and $\hat{\boldsymbol{S}}_z^{a}$ is the spin operator for the $\ha_z$-state,
defined similarly as $\hat{\boldsymbol{S}}_z$ in \rlb{Sops} with $\hc_{z,\sigma}$ replaced by $\ha_{z,\sigma}$. 

We note in passing that, by using \rlb{eff_ferro_int} and 
\begin{equation}
\hA_{x,y}^\dagger\,\hA_{x,y} = \hn_{x,\up}^{a}\hn_{x,\dn}^{a} + \hn_{y,\up}^{a}\hn_{y,\dn}^{a}
                              +\ha_{x,\up}^\dagger\ha_{x,\dn}^\dagger \ha_{y,\dn}\ha_{y,\up}
        		      +\ha_{y,\up}^\dagger\ha_{y,\dn}^\dagger \ha_{x,\dn}\ha_{x,\up},
\end{equation}
one can write the effective interaction as
\jour{
\begin{equation}
\hat{H}_\mathrm{eff}
=\cPa\Bigl\{
U^{\prime\prime}\sum_{x\in\cM} \hn_{x,\up}^{a}\hn_{x,\dn}^{a}+W\sum_{(x,y)\in\cE}
 \ha_{x,\up}^\dagger\ha_{x,\dn}^\dagger \ha_{y,\dn}\ha_{y,\up}
-J\sum_{(x,y)\in\cB}(\hat{\boldsymbol{S}}_x^a\cdot\hat{\boldsymbol{S}}_y^a-\frac{1}{4}\hn_x^a\hn_y^a)\Bigr\}\cPa,
\lb{HefftJ}
\end{equation}
}
{
\begin{eqnarray}
\hat{H}_\mathrm{eff}
&=&\cPa\Bigl\{
U^{\prime\prime}\sum_{x\in\cM} \hn_{x,\up}^{a}\hn_{x,\dn}^{a}+W\sum_{(x,y)\in\cE}
 \ha_{x,\up}^\dagger\ha_{x,\dn}^\dagger \ha_{y,\dn}\ha_{y,\up}
\nonumber\\
&&\hspace*{4cm}
-J\sum_{(x,y)\in\cB}(\hat{\boldsymbol{S}}_x^a\cdot\hat{\boldsymbol{S}}_y^a-\frac{1}{4}\hn_x^a\hn_y^a)\Bigr\}\cPa,
\lb{HefftJ}
\end{eqnarray}
}
where $U^{\prime\prime}=U^\prime+\zeta W$ and $J=2W$.
Although we won't make explicit use of this expression, it shows that our model is equivalent to the 
Hubbard model with pair-hopping and ferromagnetic exchange interaction 
defined for the $a$-band electrons\footnote{
By forbidding the double occupancy of $\ha$-states,
the model is further reduced to the ferromagnetic $t$-$J$ model, whose Hamiltonian
is given by the sum of the kinetic term $t\sum_{(x,y)\in\cE,\sigma=\up,\dn}\had_{x,\sigma}\ha_{y,\sigma}$ 
and the term proportional to  $J$ in the effective interaction \rlb{HefftJ}.
Our model in the limits $s\up\infty$ and $U\up\infty$ is thus equivalent 
to the ferromagnetic $t$-$J$ model with $J=\infty$.
}.
We also note 
that the Hubbard model with nearest neighbor interactions, 
including $\hat{H}_\mathrm{eff}$ as a special case, is studied at half-filling 
in \cite{StrackVollhardt94} and \cite{StrackVollhardt95}.
The expression \rlb{HefftJ} and the method used there 
may be helpful to estimate values of $U$ and $s$ for the occurrence of ferromagnetism
in the model stated in footnote~\ref{footnote:modela}.

\para{Condition for a finite energy state}
Let us examine the implication of the condition \rlb{Heff0}.
Note first that each term in the expression \rlb{Heff} for $\hat{H}_{\rm eff}$ is nonnegative, i.e., $\hn_{x,\up}^a\hn_{x,\dn}^a=(\ha_{x,\dn}\ha_{x,\up})^\dagger(\ha_{x,\dn}\ha_{x,\up})\ge 0$, $\hA_{x,y}^\dagger\,\hA_{x,y}\ge0$, and $\hB_{x,y}^\dagger\,\hB_{x,y}\ge0$.
One thus finds that the condition \rlb{Heff0} is satisfied if and only if
\begin{equation}
 \ha_{x,\dn}\ha_{x,\up} \kPhi = 0\quad\text{for any $x\in\cM$},
\lb{aaP}
\end{equation} 
and
\begin{equation}
 \hA_{x,y} \kPhi = 0\quad\text{and}\quad\hB_{x,y} \kPhi = 0\quad\mbox{for any $(x,y)\in\cB$}.
\end{equation}
 
From the condition \rlb{aaP}, we find $\psi(C_\up,C_\dn)=0$ for $C_\up\cap C_\dn\ne\emptyset$.
This means that the expansion \rlb{exp1} can be rearranged as 
\begin{equation}
\kPhi=\sum_{\Cs}\psi_\Cs\ket{\Psi_\Cs},
\lb{PhiPsiCs}
\end{equation}
with
\begin{equation}
\ket{\Psi_\Cs}=\Bigl(\prod_{x\in C}\had_{x,\sigma(x)}\Bigr)\vac,
\lb{PsiCs}
\end{equation}
where $C$ is a subset of $\cM$ such that $\nM=\Ne$, and $\bssigma=(\sigma(x))_{x\in C}$ 
is a spin configuration on $C$, where $\sigma(x)=\up,\dn$.
One can easily check that the state of this form satisfies $\hA_{x,y}\kPhi=0$ for any $(x,y)\in\cB$.

The expression \rlb{eff_ferro_int} suggests that the remaining condition $\hB_{x,y}\kPhi=0$ for any $(x,y)\in\cB$ is related to ferromagnetism.
The condition indeed implies that the spins of electrons on a connected component of $C$ in the expansion \rlb{PhiPsiCs} 
are coupled ferromagnetically, and have the maximum possible total spin.

More precisely, we recall the definition \rlb{Bdef}, and write the condition $\hB_{x,y}\kPhi=0$ explicitly as
\begin{equation}
(\ha_{x,\dn}\ha_{y,\up} - \ha_{x,\up}\ha_{y,\dn})
\sum_{\Cs}\psi_\Cs\Bigl(\prod_{x'\in C}\had_{x',\sigma(x')}\Bigr)\vac=0.
\end{equation}
Inspection shows that this is satisfied when one has
\begin{equation}
\psi_\Cs=\psi_{C,\bssigma_{x\leftrightarrow y}},
\lb{exch}
\end{equation}
for any $x,y\in C$ such that $(x,y)\in\cB$.
Here $\bssigma_{x\leftrightarrow y}$ is the configuration 
obtained by switching $\sigma(x)$ and $\sigma(y)$ in the original configuration $\bssigma$.
The condition \rlb{exch} implies the above claim that the electrons on a connected component of $C$ must be coupled ferromagnetically.
We have derived the existence of strong ferromagnetic coupling in finite energy states.

\para{Ferromagnetism in finite energy states}

One might probably feel that the above derivation of ferromagnetic coupling in connected components of $C$ is almost a goal for us.
This is however far from the case.
If the dimension is higher than one\footnote{
In one dimension, the situation is totally different.
One expects, and can indeed prove, that the present model exhibits metallic ferromagnetism for any $0<\Ne\le\nM$.
See, e.g., \cite{Tasaki98a}.
}, and the electron density is sufficiently low, it is expected that the electrons behave as ``interacting waves'' in which electrons avoid each other without causing too much energy loss. 
Consequently the ground state should exhibit paramagnetism.
Although it may be extremely difficult to prove this fact, it is not too difficult to prove, in the line of \cite{Pieri}, that the present model in $d\ge3$ cannot have ground state with $\Stot=\Smax$ when the electron number is sufficiently small (see also Theorem~3.3 of \cite{Tasaki98a}).

If the density $\Ne/\nM<1$ is sufficiently close to 1, on the other hand,  the set $C$ in the expansion \rlb{PhiPsiCs} of the ground state is expected to have a large connected component.
Then the ground state should exhibit metallic ferromagnetism.
We shall make this idea concrete to prove our theorems.

Let us treat the simplest case, and prove Theorems~\ref{t:Feasy0} and \ref{t:Feasy1}.
Suppose that the electron number $\Ne$ satisfies $\nM\ge\Ne>\nM-\zeta$, i.e., the number of the ``holes'' is less than the coordination number $\zeta$.
Take an arbitrary finite energy state $\kPhi$, and consider the expansion \rlb{PhiPsiCs}.
Any $C$ in the expansion is connected because $|C|=\Ne>\nM-\zeta$.
Thus all the electrons are coupled ferromagnetically, and $\kPhi$ has $\Stot=\Smax$.

%%%%%%%%%%%%%%%%%%%%%%%%%%%%%%%%%%%%%%%%%%
\section{Bound for the energy expectation value}
\label{s:E}
To have a stronger control of the ground state, we shall derive a lower bound for the energy expectation value.
Let $\kPhi$ be an arbitrary state which satisfies the finite energy condition \rlb{fe}.
Then one has
\begin{equation}
\EP:=\bra{\Phi}\Ham\kPhi=
\bra{\Phi}
\,t\sumtwo{(x,y)\in\cE}{\sigma=\up,\dn}\had_{x,\sigma}\ha_{y,\sigma}
\kPhi
\lb{EP1}
\end{equation}
because of \rlb{fe2} and \rlb{ccP}.
By using the expansion \rlb{PhiPsiCs}, we can rewrite this as
\begin{equation}
\EP=t\sumtwo{(x,y)\in\cE}{\sigma=\up,\dn}\ \sum_{\Cs,C',\bssigma'}
(\psi_{C',\bssigma'})^*\,\psi_{\Cs}
\bra{\Psi_{C',\bssigma'}}\had_{x,\sigma}\ha_{y,\sigma}\ket{\Psi_{\Cs}}.
\lb{EP2}
\end{equation}
Note that $\bra{\Psi_{C',\bssigma'}}\had_{x,\sigma}\ha_{y,\sigma}\ket{\Psi_{\Cs}}$ can only be 0 or $\pm1$ because of the anticommutation relation \rlb{aa} and the definition \rlb{PsiCs}.
Also note that, for a given combination of $C$, $\bssigma$, $C'$, and $\bssigma'$, the quantity $\bra{\Psi_{C',\bssigma'}}\had_{x,\sigma}\ha_{y,\sigma}\ket{\Psi_{\Cs}}$ is nonvanishing at most for a single combination of $(x,y)$ and $\sigma$.
We thus see that
\begin{align}
\chi(C',\bssigma';\Cs)&:=\sumtwo{(x,y)\in\cE}{\sigma=\up,\dn}
\bigl|\bra{\Psi_{C',\bssigma'}}\had_{x,\sigma}\ha_{y,\sigma}\ket{\Psi_{\Cs}}\bigr|
\nl
&=
\begin{cases}
1&\text{if $(C,\bssigma)$ changes to $(C',\bssigma')$ by moving one electron};\\
0&\text{otherwise}.
\end{cases}
\lb{chi}
\end{align}
Then by using the trivial inequality
\begin{equation}
\pm\bigl\{
(\psi_{C',\bssigma'})^*\,\psi_{\Cs}+(\psi_{\Cs})^*\psi_{C',\bssigma'}
\bigr\}
\ge-|\psi_{\Cs}|^2-|\psi_{C',\bssigma'}|^2,
\end{equation}
we can bound \rlb{EP2} as
\begin{align}
\EP&=t\sum_{\Cs,C',\bssigma'}\pm\chi(C',\bssigma';\Cs)\,
(\psi_{C',\bssigma'})^*\,\psi_{\Cs}
\nl
&\ge-\frac{t}{2}\sum_{\Cs,C',\bssigma'}\chi(C',\bssigma';\Cs)
\bigl\{|\psi_{\Cs}|^2+|\psi_{C',\bssigma'}|^2\bigr\}
\nl
&=-t\sum_{\Cs}\gamma(C)\,|\psi_{\Cs}|^2,
\lb{EP3}
\end{align}
where we used the symmetry $\chi(C',\bssigma';\Cs)=\chi(\Cs;C',\bssigma')$.
We have defined
\begin{equation}
\gamma(C):=\sum_{C',\bssigma'}\chi(\Cs;C',\bssigma'),
\lb{gdef}
\end{equation}
where the sum does not depend on $\bssigma$ because of the symmetry.
Note that $\gamma(C)$ is the number of possible ``hops'' of electrons which can take place in the configuration $C$.
Since we are interested in the situation where the number of electrons $\Ne$ is close to (but less than) $\nM$, the number $\gamma(C)$ is mainly determined by the location of  ``holes'', i.e., $x\in\cM$ such that $x\not\in C$.

The rest of our analysis is based on the bound \rlb{EP3}.

%%%%%%%%%%%%%%%%%%%%%%%%%%%%%%%%%%%%%%%%%%
\section{Saturated ferromagnetism in the triangular lattice model}
\label{s:SFt}

\begin{figure}[btp]
\begin{center}
\jour{\includegraphics[width=6cm]{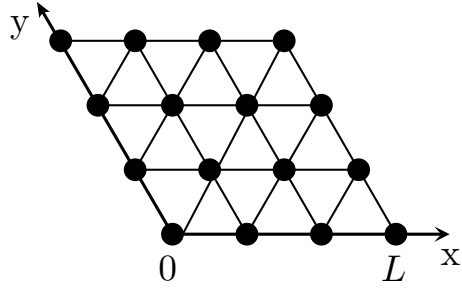}\vspace{-5mm}}
{\includegraphics[width=5cm]{tri.eps}}
\end{center}
\caption[dummy]{
The triangular lattice with $3\times3=9$ sites.
We impose periodic boundary conditions in the x and the y directions, identifying 0 and $L$.
}
\label{fig:tri}
\end{figure}

In the present and the next sections, we shall concentrate on the simplest nontrivial case where $\cM$ is the triangular lattice.
To be precise we let $\cM$ be the triangular lattice with $L^2$ sites with periodic boundary conditions as depicted in Fig.~\ref{fig:tri}.
The orientations of bonds are arbitrary.

%%%%%%%%%%%%%%%%%%%%%%%%%%%%%%%%%%%%%%
%%%%%%%%%%%%%%%%%%%%%%%%%%%%%%%%%%%%%%

We shall prove the following theorem, which is the two-dimensional version of Theorem~\ref{t:SF}.

%%%%%%%%%%%%%%%%%%%%%%
\begin{theorem}\label{t:SFt}
Suppose that the electron number $\Ne$ satisfies
\begin{equation}
\nM\ge\Ne\ge\nM-\frac{1}{2}\nM^{1/2}.
\end{equation}
Then, in the limits $U\up\infty$ and $s\up\infty$, ground states of the model have $\Stot=\Smax$.
\end{theorem}
%%%%%%%%%%%%%%%%%%%%%%

Let $\kPhi$ be an arbitrary state satisfying the finite energy condition \rlb{fe}, and expand it as \rlb{PhiPsiCs}.
Let $\Nh:=\nM-\Ne=\nM-|C|$ be the number of ``holes''.
We shall derive an upper bound in terms of $\Nh$ for $\gamma(C)$ defined by \rlb{gdef} .

%%%%%%%%%%%%%%%%%%%%%%%%%%%%%%%%%%%%%%
\begin{figure}[btp]
\begin{center}
\jour{\includegraphics[width=3cm]{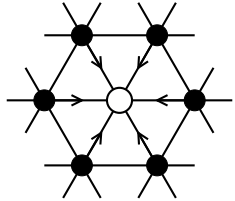}\vspace{-8mm}}
{\includegraphics[width=3cm]{single.eps}}
\end{center}
\caption[dummy]{
There are six possible hops associated with a single isolated hole.
}
\label{fig:single}
\end{figure}
%%%%%%%%%%%%%%%%%%%%%%%%%%%%%%%%%%%%%%

Suppose that, in the configuration $C$, all the holes are isolated
in the sea of electrons as in Fig.~\ref{fig:single}.
Then, since one of the six neighboring electrons may hop into 
each hole, 
there are $6\Nh$
configurations $C'$ which contribute to the sum in \rlb{gdef}.
By noting that the number of possible hops decreases when some holes are neighboring with each other, we get
\begin{equation}
\gamma(C)\le6\Nh,
\lb{gC6}
\end{equation}
which is valid for any possible $C$.

Let us now assume that the state $\kPhi$ is a simultaneous eigenstate of $\Ham$ and $(\bStot)^2$ which does not exhibit saturated ferromagnetism, i.e., it has the total spin $\Stot<\Smax$.
Then from the discussion in section~\ref{s:tJ} we see that any configuration $C$ contributing to the expansion \rlb{PhiPsiCs} is not connected.

%%%%%%%%%%%%%%%%%%%%%%%%%%%%%%%%%%%%%%
\begin{figure}[btp]
\begin{center}
\jour{\includegraphics[width=6cm]{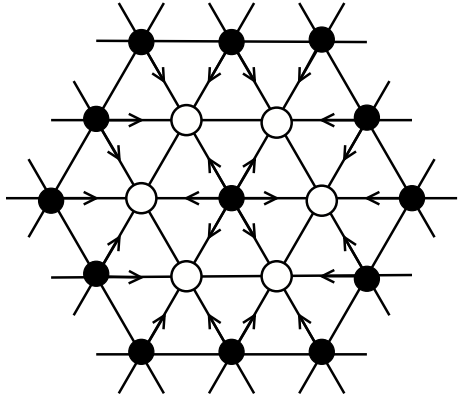}\vspace{-5mm}}
{\includegraphics[width=5cm]{six.eps}}
\end{center}
\caption[dummy]{
Six holes are surrounding a single electron, thus making $C$ non-connected.
There are twenty four possible hops in this configuration.
}
\label{fig:six}
\end{figure}
%%%%%%%%%%%%%%%%%%%%%%%%%%%%%%%%%%%%%%

When $C$ is not connected, not all the holes in $C$ can be isolated.
Consequently the upper bound for $\gamma(C)$ is reduced from \rlb{gC6}.
Inspection shows that non-connected $C$ with maximum $\gamma(C)$ has six holes surrounding a single electron as in Fig.~\ref{fig:six}, and $\Nh-6$ isolated holes.
Since there are $4\times6=24$
possible hops associated with the ring of six holes, we find that
\begin{equation}
\gamma(C)\le 6(\Nh-6) + 24=6\Nh-12,
\lb{gC1}
\end{equation}
whenever $C$ is not connected.
Substituting this into \rlb{EP3}, and noting that $\sum_{\Cs}|\psi_{\Cs}|^2=1$, we find that the energy expectation $\EP$ of $\kPhi$ satisfies
\begin{equation}
\EP\ge-t\max_C\gamma(C)=-6t\Nh+12t.
\lb{EPL1}
\end{equation}

We shall compare the lower bound \rlb{EPL1} with
the lowest energy among the states (with the same $\Ne$) with the maximum possible spin, i.e., $\Stot=\Smax$.
This is an easy task for the Hubbard model since states with $\Stot=\Smax$ (which can always be represented only by using up-spin electrons) are not affected by the interaction.
The problem reduces to that of non-interacting spinless electrons.

Let the single-electron energy eigenvalues $\ep_1,\ldots,\ep_{\nM}$ be the solution of the single-electron Schr\"{o}dinger equation \rlb{sSch}, where we assume $\ep_j\le\ep_{j+1}$.
Then the lowest energy among the states with $\Stot=\Smax$ is
\begin{equation}
\Ef=\sum_{j=1}^{\Ne}\ep_j=-\sum_{j=\Ne+1}^{\nM}\ep_j,
\lb{Ef0}
\end{equation}
where we noted that\footnote{
$\ep_1,\ldots,\ep_{\nM}$ are the eigenvalues of the matrix $\mathsf{T}=(t_{x,y})_{x,y\in\cM}$, with $t_{x,y}=t$ if $(x,y)\in\cE$ and $t_{x,y}=0$ otherwise.
Then $\operatorname{Tr}[\mathsf{T}]=0$ implies $\sum_{j=1}^{\nM}\ep_j=0$.
} $\sum_{j=1}^{\nM}\ep_j=0$.
Now explicit computation shows that the solution of \rlb{sSch} for the triangular lattice is
\begin{equation}
\ep(\bsk)=2t\bigl\{\cos\kx+\cos\ky+\cos(\kx+\ky)\bigr\},
\lb{epk}
\end{equation}
where $\bsk=(\kx,\ky)$ with $k_\alpha=(2\pi/L)n_\alpha$, $n_\alpha=0,\pm1,\ldots,\pm(L-1)/2$ for $\alpha=\mathrm{x},\mathrm{y}$.
We thus find 
\begin{equation}
\Ef=\mintwo{\calK}{(|\calK|=\Nh)}\left(-\sum_{\bsk\in\calK}\ep(\bsk)\right),
\end{equation}
where $\calK$ is an arbitrary set of $\bsk$'s with $\Nh$ elements.
By using $\cos x\ge1-x^2/2$, we find
\begin{align}
\Ef&\le\mintwo{\calK}{(|\calK|=\Nh)}\left(\sum_{\bsk\in\calK}\Bigl[
-6t+t\bigl\{\kx^2+\ky^2+(\kx+\ky)^2\bigr\}
\Bigr]\right)
\nl
&\le-6t\Nh+\mintwo{\calK}{(|\calK|=\Nh)}\left(3t\sum_{\bsk\in\calK}|\bsk|^2\right),
\lb{Ef2}
\end{align}
where we used $2\kx\ky\le\kx^2+\ky^2$ to get the final bound.
We can solve the minimization problem by replacing the sum with the integral as\footnote{
To make this estimate into a rigorous bound is a routine work, which we shall omit here.
}
\begin{equation}
\mintwo{\calK}{(|\calK|=\Nh)}\left(\sum_{\bsk\in\calK}|\bsk|^2\right)
\simeq\left(\frac{L}{2\pi}\right)^2\int_{|\bsk|\le k_0}d^2\bsk\,|\bsk|^2
=2\pi\nM\left(\frac{\Nh}{\nM}\right)^2,
\end{equation}
where we determined $k_0$ as $k_0=\sqrt{4\pi\Nh/\nM}$ from the condition $\Nh\simeq(L/2\pi)^2\int_{|\bsk|\le k_0}d^2\bsk$, noting that $L^2=\nM$.
We thus get the upper bound
\begin{equation}
\Ef\le-6t\Nh+6\pi t\nM\left(\frac{\Nh}{\nM}\right)^2.
\lb{EfU}
\end{equation}

One can say that the ground states of the model have $\Stot=\Smax$ if one has $\EP>\Ef$.
By using the lower bound \rlb{EPL1} and the upper bound \rlb{EfU}, we find that $\EP>\Ef$ is guaranteed if
\begin{equation}
12t>6\pi t\nM\left(\frac{\Nh}{\nM}\right)^2,
\end{equation}
which is equivalent to $\Nh<\sqrt{(2/\pi)\nM}$.
In Theorem~\ref{t:SFt}, we used a sufficient condition $\Nh\le\sqrt{\nM}/2$.

%%%%%%%%%%%%%%%%%%%%%%%%%%%%%%%%%%%%%%%%%%
\section{Ferromagnetism in the triangular lattice model}
\label{s:Ft}

Theorem~\ref{t:SFt} only allows the number of the holes $\Nh$ to be proportional to $\sqrt{\nM}$.
Here we shall relax the characterization of ferromagnetism, and prove the following theorem which allows much larger $\Nh$.
This is the two dimensional version of Theorem~\ref{t:F}.

%%%%%%%%%%%%%%%%%%%%%%
\begin{theorem}\label{t:Ft}
Fix an arbitrary constant $\nu$ such that $0<\nu<1$.
Suppose that the electron number $\Ne$ satisfies
\begin{equation}
\nM\ge\Ne\ge\nM-\frac{(1-\nu)^{1/4}}{5}\nM^{3/4}.
\end{equation}
Then, in the limits $U\up\infty$ and $s\up\infty$,  ground states of the model have 
$\Stot>\nu\Smax$.
\end{theorem}
%%%%%%%%%%%%%%%%%%%%%%

Take an arbitrary $\kPhi$ which satisfies the finite energy condition \rlb{fe}, and expand it as \rlb{PhiPsiCs}.
Take $C\subset\cM$ which corresponds to a nonvanishing $\psi_{\Cs}$.
We decompose $C$ into connected components as $C=C_0\cup C_1\cup\cdots\cup C_n$, where $C_0$ is chosen so that $|C_0|\ge|C_\ell|$ for any $\ell\ne0$.
The following elementary lemma is essential.

\begin{lemma}
Take a constant $\nu\in(0,1)$.
Suppose that a finite energy state $\kPhi$ is an eigenstate of $(\bStot)^2$ with $\Stot\le\nu\Smax$.
Then for any $C$ such that $\psi_{\Cs}\ne0$ in the expansion \rlb{PhiPsiCs}, one has
\begin{equation}
\sum_{\ell=1}^n|C_\ell|\ge\frac{\Ne}{2}(1-\nu).
\lb{Cell>}
\end{equation}
\end{lemma}

\noindent
{\em Proof\/:}
One has either $\sum_{\ell=1}^n|C_\ell|\ge\Ne/2$ or $\sum_{\ell=1}^n|C_\ell|<\Ne/2$.
Since \rlb{Cell>} automatically holds for the former, assume that the latter is the case.
We then have $|C_0|\ge\Ne/2$.
We claim that the total spin of $\kPhi$ must satisfy
\begin{equation}
\Stot\ge\frac{1}{2}\bigl\{|C_0|-\sum_{\ell=1}^n|C_\ell|\bigr\}
=\frac{\Ne}{2}-\sum_{\ell=1}^n|C_\ell|.
\lb{Stot>}
\end{equation}
This is because the minimum $\Stot$ is realized by first coupling all the spins on $C_1,\ldots,C_n$ to have the maximum spin, and then coupling this with spin state on $C_0$ to minimize the total spin.
By using the assumption $\Stot\le\nu\Smax$, we get the desired \rlb{Cell>} from \rlb{Stot>}.\quad\qedm

\bigskip

As in the lemma, we assume that a finite energy state $\kPhi$ is an eigenstate of $(\bStot)^2$ with $\Stot\le\nu\Smax$, where $\nu\in(0,1)$.
We know from the lemma that, in a configuration $C$ with $\psi_{\Cs}\ne0$, there are $\Ne'$ electrons which do not belong to the largest connected component $C_0$, where $\Ne'\ge(\Ne/2)(1-\nu)$.
By using this information, we shall show (at the end of the present section) that $\gamma(C)$ defined in \rlb{gdef} is bounded as
\begin{equation}
\gamma(C)< 6\Nh-\sqrt{2\Ne'}\le 6\Nh-\sqrt{(1-\nu)\Ne}.
\lb{gC0}
\end{equation}

Then the bound \rlb{EP3}, along with $\sum_{\Cs}|\psi_{\Cs}|^2=1$, implies that the energy expectation of the state $\kPhi$ satisfies 
\begin{equation}
\EP\ge-t\max_C\gamma(C)
>
-6t\Nh+t\sqrt{(1-\nu)\Ne}
>-6t\Nh+
\frac{\pi t}{4}
\sqrt{(1-\nu)\nM},
\lb{EPL2}
\end{equation}
where we assumed $\Ne>(\pi/4)^2\nM$, which is harmless since our result is valid when $\Ne$ is close to $\nM$.
The factor $(\pi/4)^2$ is chosen to make the final result simple.

Again let $\Ef$ be the minimum energy among the states with $\Stot=\Smax$.
If we have $\EP>\Ef$ for any $\kPhi$ whose total spin satisfies $\Stot\le\nu\Smax$, we find that the ground state must have $\Stot>\nu\Smax$.

By comparing the upper bound \rlb{EfU} for $\Ef$ and the lower bound \rlb{EPL2}, we see that $\EP>\Ef$ is guaranteed if
\begin{equation}
\frac{\pi t}{4}
\sqrt{(1-\nu)\nM}\ge6\pi t\nM\left(\frac{\Nh}{\nM}\right)^2,
\end{equation}
or
\begin{equation}
\Nh\le\frac{(1-\nu)^{1/4}}{5}
\nM^{3/4}.
\end{equation}

It remains to prove the upper bound \rlb{gC0} under the assumption that there are $\Ne'$ electrons which do not belong to the largest connected component $C_0$.
We note that
there are two types of configurations in which $\Ne'$ electrons are separated from the others.
In one type, there are two loops formed by holes which lap around the periodic lattice in the x (or y) direction.
In the other type, there are loops formed by holes which surround $\Ne'$ electrons.  

We first focus on  configurations of the former type. 
The maximum $\gamma(C)$ is obtained when the loops are straight line with length $L$, and all the remaining $\Nh-2L$ holes are isolated.
Since there are four possible hops for each hole in the loops and six possible hops for the isolated holes, 
we have the upper bound
\begin{equation}
 \gamma(C) \le 6(\Nh-2L)+4\cdot2L=6\Nh-4L<6\Nh-4\sqrt{2\Ne'},
 \lb{gc11}
\end{equation}
where we used $\Ne'< L^2/2$.

We next examine the second type (which is more normal).
In a configuration $C$ of this type, we suppose that there are $\Nh'$ holes which form the loops.
Then we have
\begin{equation}
 \gamma(C)\le 6(\Nh-\Nh')+4\Nh'=6\Nh-2\Nh'
\end{equation}
as in \rlb{gc11}.
This implies that we can obtain the upper bound by considering a configuration 
with a possible minimum value of $\Nh'$. 
Apparently, the minimum value is attained by a configuration with a single loop.
Furthermore, for configurations with $\Nh'>2L$, the upper bound is given by \rlb{gc11}.  
In the following, we estimate the upper bound by considering configurations with 
a single loop of holes less than $2L$.

\begin{figure}[btp]
\begin{center}
\jour{\includegraphics[width=6cm]{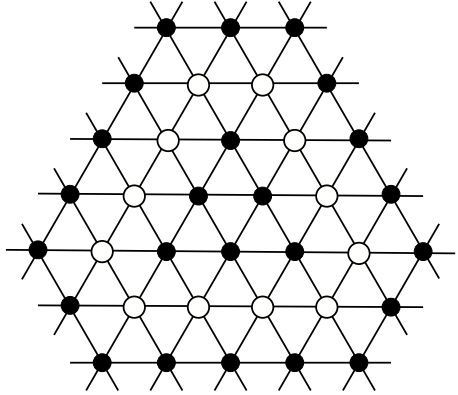}\vspace{-5mm}}
{\includegraphics[width=5.5cm]{RT.eps}}
\end{center}
\caption[dummy]{
The configuration which maximizes $\gamma(C)$ when $m=3$.
}
\label{fig:RT}
\end{figure}

To begin with, let us consider a simple case where $\Ne'$ can be written as $\Ne'=m(m+1)/2$ with some $m=1,2,\ldots$, and $3(m+1)<2 L$.  
Then the maximum value of $\gamma(C)$ is attained when $\Ne'$ electrons form an equilateral triangle surrounded by $3(m+1)$ holes, and all other holes are isolated
\footnote{
In the configurations obtained by moving electrons from sites at the apexes 
to sites touching sides of the triangle, $\Ne'$ electrons are surrounded by $3(m+1)$ holes.
These configurations also attain the maximum.
}.
See Fig.~\ref{fig:RT}.
Since four hops are possible for each 
hole surrounding the $\Ne'$ electrons, there correspond $12(m+1)$ hops to the equilateral triangle.
As in \rlb{gC1}, we can bound 
the total number of hops as
\begin{equation}
\gamma(C)\le 6\{\Nh-3(m+1)\}+12(m+1)=6\Nh-6(m+1)<6\Nh-6\sqrt{2\Ne'},
\lb{gC2}
\end{equation}
where we used $\sqrt{2\Ne'}<(m+1)$.

%%%%%%%%%%%%%%%%%%%%%%%%%%%%%%%%%%%%%%
%%%%%%%%%%%%%%%%%%%%%%%%%%%%%%%%%%%%%%

For a general $\Ne'$ which is not necessarily of the form $m(m+1)/2$, the maximum value of $\gamma(C)$ is attained by a configuration where $\Ne'$ electrons are surrounded by $\Nh'$ holes with $\Nh'<2L$.  
Denoting by $m$ the positive integer such that $m(m+1)/2\le \Ne' < (m+1)(m+2)/2$, we see that $\Nh'$ satisfies $\Nh'\ge3(m+1)$. 
We thus obtain 
\begin{equation}
\gamma(C)\le 6\Nh-6(m+1)< 6\Nh-6\sqrt{2\Ne'}+6<6\Nh-\sqrt{2\Ne'},
\lb{gNh}
\end{equation}
where we used $6/5<\sqrt{2\Ne'}<(m+2)$.

From \rlb{gc11}, \rlb{gC2}, and \rlb{gNh}, we get the desired upper bound \rlb{gC0}.

%%%%%%%%%%%%%%%%%%%%%%%%%%%%%%%%%%%%%%%%%%
\section{Models on other lattices}
\label{s:lattice}
Finally let us describe how Theorems~\ref{t:SFt} and \ref{t:Ft} are extended to other lattices, thus proving Theorems~\ref{t:SF} and \ref{t:F}.
We first recall that the results in sections~\ref{s:tJ} and \ref{s:E}, especially the essential lower bound \rlb{EP3} for the energy expectation value, are all valid for the model on any lattice.

%%%%%%%%%%%%%%%%%%%%%%%%%%%%%%%%%%%%%%
\begin{figure}[btp]
\begin{center}
\jour{\includegraphics[width=12cm]{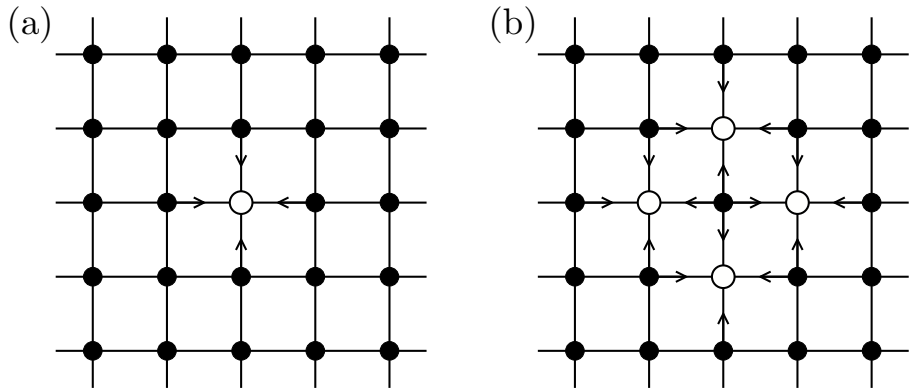}\vspace{-7mm}}
{\includegraphics[width=11cm]{square.eps}}
\end{center}
\caption[dummy]{
In the model based on the square lattice with nearest neighbor hopping, (a)~there are four hops associated with an isolated hole, and (b)~sixteen hops with an isolated electron surrounded by four holes.
There is no reduction of the number of hops.
}
\label{fig:square}
\end{figure}
%%%%%%%%%%%%%%%%%%%%%%%%%%%%%%%%%%%%%%

We note however that our proofs in sections~\ref{s:SFt} and \ref{s:Ft}, as they are, 
can not be extended
to such simple lattices as the square lattice.
To see this we examine the extension of the bounds \rlb{gC1} for $\gamma(C)$ and \rlb{EPL1} for $\EP$.
When $\cM$ is taken as the square lattice, a single isolated hole allows four hops as in Fig.~\ref{fig:square}~(a).
Thus the general bound corresponding to \rlb{gC6} is $\gamma(C)\le4\Nh$.
If $C$ is not connected and there is an isolated electron, there are still $4\times4=16$ possible hops as in Fig.~\ref{fig:square}~(b).
This means that we cannot prove an improved upper bound for $\gamma(C)$.
We cannot prove that the existence of a non-connected component in $C$ raises the energy.
We believe that this difficulty is not of fundamental nature, but comes from our simple strategy to use only the elementary lower bound \rlb{EP3}.
We expect that we can cover the case of the square lattice by improving the estimate\footnote{
Although we impose some geometric constraints to the configuration $C$, the configuration $C'$ that appears in the derivation of \rlb{EP3} is arbitrary in the current estimate.
We should be able to get better bounds by introducing restriction to $C'$ as well.
}, but we shall not go into such an extension here.

The situation is the same for higher dimensions.
Our argument does not work for models on, say, the standard hypercubic lattice.

In order for a straightforward extension of our proof to work, the lattice should have dimension larger than one, and satisfy the condition that 
the set $S:=\{y\in\cM\,|\,(x_0,y)\in\cE\}$ (for any fixed $x_0\in\cM$)
is connected via bonds in $\cE$.
We say that the lattice $\cM$ is ``closely packed'' when this condition is satisfied.
As we noted below Theorem~\ref{t:Feasy1}, the triangular and the checkerboard lattices (where bonds are identified with nearest neighbor pairs of sites) are examples.
We can also consider lattices whose bonds consist not only of nearest neighbor paris of sites but also of extra non-nearest neighbor pairs to construct ``closely packed lattices''.

%%%%%%%%%%%%%%%%%%%%%%%%%%%%%%%%%%%%%%
\begin{figure}[btp]
\begin{center}
\jour{\includegraphics[width=4cm]{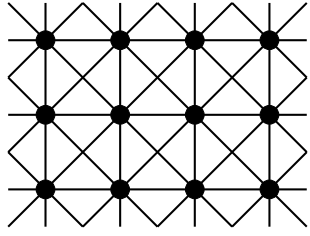}\vspace{-5mm}}
{\includegraphics[width=4cm]{extra.eps}}
\end{center}
\caption[dummy]{
The square lattice with next-nearest-neighbor hoppings.
Our proof works when $\cM$ is taken as this lattice, and Theorems~\ref{t:SF} and \ref{t:F} are valid.
}
\label{fig:extra}
\end{figure}
%%%%%%%%%%%%%%%%%%%%%%%%%%%%%%%%%%%%%%

Let $\cM$ be a uniform $d$-dimensional lattice with coordination number $\zeta$ which 
is closely packed, i.e., satisfies the above condition.
Then we can prove Theorems~\ref{t:SF} and \ref{t:F}.
Let us briefly describe the difference from the case of the triangular lattice.

We start from the estimate of $\Ef$.
We assume that the single-electron energy eigenvalues of the model satisfy
\begin{equation}
\ep(\bsk)\ge\zeta t-C_1t|\bsk|^2,
\end{equation}
with a constant $C_1$ which depends only on the lattice structure and not on the lattice size.
Then, exactly as in \rlb{EfU}, we obtain
\begin{equation}
\Ef\le-\zeta t \Nh+C_2t\nM\left(\frac{\Nh}{\nM}\right)^{(d+2)/d},
\lb{Efg}
\end{equation}
with a constant $C_2$.

To prove Theorem~\ref{t:SF}, we note that the condition on the lattice stated above implies
\begin{equation}
\gamma(C)\le\zeta\Nh-C_3,
\end{equation}
for any non-connected $C$ as in \rlb{gC1}, with a constant $C_3$.
This implies the lower bound
\begin{equation}
\EP\ge-\zeta t\Nh+C_3t,
\end{equation}
for any $\kPhi$ with $\Stot<\Smax$.
Combining with the upper bound \rlb{Efg}, we get Theorem~\ref{t:SF}.

To prove an upper bound corresponding to \rlb{gC0}, note that we need at least $(\text{const.})(\Ne')^{(d-1)/d}$ holes to isolate $\Ne'$ electrons from the ``sea'' of the remaining electrons.
Thus, as in \rlb{gC0}, we can show that 
\begin{equation}
 \gamma(C)\le\zeta\Nh-(\text{const.})(\Ne')^{(d-1)/d}\le \zeta\Nh-(\text{const.})\left\{(1-\nu)\Ne\right\}^{(d-1)/d}.
\end{equation}
Then, corresponding to \rlb{EPL2}, we can show the bound
\begin{equation}
\EP>-\zeta t\Nh+C_4t\bigl\{(1-\nu)\nM\bigr\}^{(d-1)/d}.
\end{equation}
Again combining with the upper bound \rlb{Efg}, we get Theorem~\ref{t:F}.

\jour{
\bigskip
{\small It is a pleasure to thank Hosho Katsura for valuable discussions.
The present work was supported by JSPS Grants-in-Aid for Scientific Research no.~25400407.}
}
{
\begin{acknowledgements}
It is a pleasure to thank Hosho Katsura for valuable discussions.
The present work was supported by JSPS Grants-in-Aid for Scientific Research no.~25400407.
\end{acknowledgements}
}

%\newpage

\end{document}